\documentclass[twocolumn, prb,showpacs,superscriptaddress]{revtex4-1}

\usepackage{amsmath,amssymb}
\usepackage{amsmath}
\usepackage{graphicx}
\usepackage{subfigure}
\usepackage{color}
\newcommand{\Ham}{\ensuremath{\mathcal{H}}}
\newcommand{\dg}{\dagger}
\newcommand{\pd}{{\phantom{\dagger}}}
\newcommand{\uar}{\uparrow}
\newcommand{\dar}{\downarrow}

\newcommand{\ket}[1]{|{#1}\rangle}
\newcommand{\braket}[1]{\langle{#1}\rangle}

\begin{document}
\title{Strongly bound yet light bipolarons for  double-well
  electron-phonon coupling}

\author{Clemens P.J. \surname{Adolphs}}
\email{cadolphs@phas.ubc.ca}
\affiliation{\!Department \!of \!Physics and Astronomy, \!University of\!
  British Columbia, \!Vancouver, British \!Columbia,\! Canada,\! V6T \!1Z1}

\author{Mona Berciu}
\affiliation{\!Department \!of \!Physics and Astronomy, \!University of\!
  British Columbia, \!Vancouver, British \!Columbia,\! Canada,\! V6T \!1Z1}
\affiliation{\!Quantum Matter \!Institute, \!University of British Columbia,
  \!Vancouver, British \!Columbia, \!Canada, \!V6T \!1Z4}

\date{\today}

\begin{abstract}
  We use the Momentum Average approximation (MA) to study the
  ground-state properties of strongly bound bipolarons in the
  double-well electron-phonon (el-ph) coupling model, which 
  describes certain intercalated lattices where the linear term in the
  el-ph coupling vanishes due to symmetry.  We show that this model
  predicts the existence of strongly bound yet lightweight bipolarons
  in some regions of the parameter space. This provides a novel
  mechanism for the appearance of such bipolarons, in addition to
  long-range el-ph coupling and special lattice geometries.
\end{abstract}

%63.20.kd Phonon-electron interactions
%63.20.Ry Anharmonic lattice modes
%71.38.-k 	Polarons and electron-phonon interactions
%71.38.Ht 	Self-trapped or small polarons
%TODO pacs numbers same as single-polaron paper...?

\maketitle

\section{Introduction}
The coupling of charge carriers to lattice degrees of freedom
(phonons) plays an important role in determining the properties of a
wide range of materials such as organic
semiconductors,\cite{organic_1, organic_2} cuprates,\cite{cuprates_1,
  cuprates_2, cuprates_3, cuprates_4, cuprates_5, cuprates_6}
manganites,\cite{manganites} two-gap superconductors like
MgB$_2$,\cite{mgb_1, mgb_2, mgb_3, mgb_4} and many more.

When a charge carrier becomes dressed by a cloud of phonons, the
quasi-particle that forms -- the polaron -- may have quite different
properties from the free particle, such as a larger effective mass and
renormalized interactions with other particles.  One particularly
interesting effect of the latter is the formation of bipolarons, where
an effective attraction mediated by exchange of phonons binds the
carriers together. If the binding is strong enough, the two phonon clouds
merge into one, resulting in a so-called S0 bipolaron. Weaker
binding, where each polaron maintains its cloud and the binding is
mediated by virtual visits to the other carrier's cloud, is also
possible and results in a S1 bipolaron.\cite{bonca_bp, macridin}

The existence of bipolarons is interesting for many reasons. For
instance, it has been suggested that Bose-Einstein condensation of
bipolarons might be responsible for superconductivity in some
high-$T_c$ materials.\cite{*[{See }] [{ for an overview.}]
  alexandrov_review}  For this to occur, the bipolaron must be
strongly bound so it can survive up to high temperatures. However,
such strong binding generally requires strong electron-phonon
coupling. In most simple models of el-ph coupling such as the
Holstein model,\cite{holstein}  this also
results in a large effective mass of the
bipolaron\cite{bonca_bp, macridin} which severely reduces its
mobility and makes it likely to become localized by even
small amounts of disorder.

For this reason, much of the theoretical work on bipolarons is focused
on finding models and parameter regimes for which the bipolaron is
strongly bound yet relatively light. So far, successful mechanism are
based either on longer-range electron-phonon
interactions\cite{extended_HH,hague_superlight,alexandrov_superlight_2012,daven}
or on special lattice geometries such as one-dimensional ladders or
triangular lattices.\cite{hague_light_bp}

Here we show that the recently proposed (short-range) double-well
el-ph coupling model\cite{double_well} also predicts the existence of
strongly bound bipolarons with relatively low effective mass in
certain regions of the parameter space, thus revealing another
possible mechanism for their appearance. Our study uses the Momentum
Average (MA) approximation,\cite{MA_berciu, MA_goodvin,
  adolphs_nonlinear, double_well} which we validate with exact
diagonalization in an enlarged variational space. 
Since in the single-particle case the dimensionality of the
  underlying lattice had little qualitative impact, we focus here on
  the one-dimensional case.

This work is organized as follows. In Section~\ref{sec:model} we
introduce the Hamiltonian for the double-well model and in
Section~\ref{sec:formalism} we discuss the methods we use to solve it. In
Section~\ref{sec:results} we present results for the bipolaron binding
energy and effective mass, and in Section~\ref{sec:conclusion} we summarize
our conclusions and an outlook for future work.

\section{Model}\label{sec:model}

The double-well el-ph coupling model was introduced in
Ref.~\onlinecite{double_well} for the single polaron case. For ease of
reference, we repeat some of its motivation and introduction here.

\begin{figure}[t]
  \centering
\subfigure[]{\includegraphics[width=0.4\textwidth]{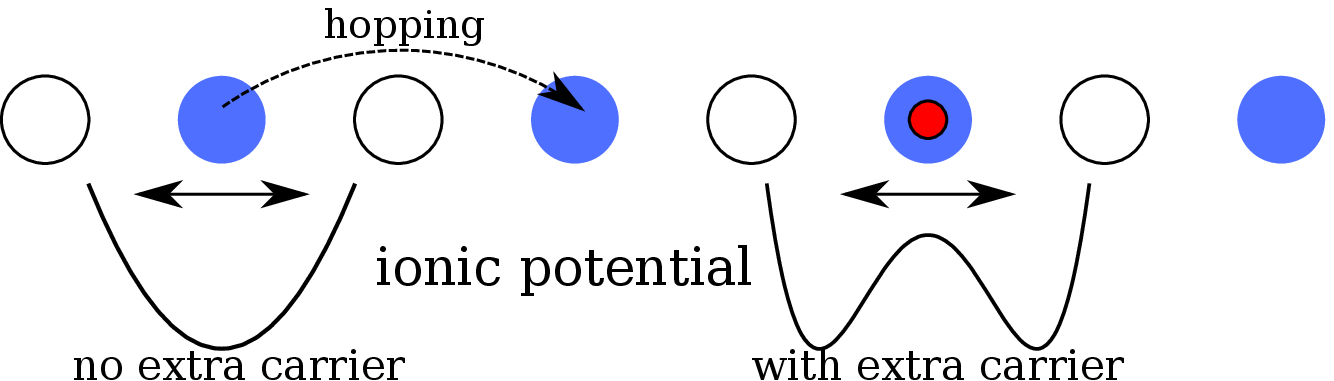}}
\subfigure[]{\includegraphics[width=0.2\textwidth]{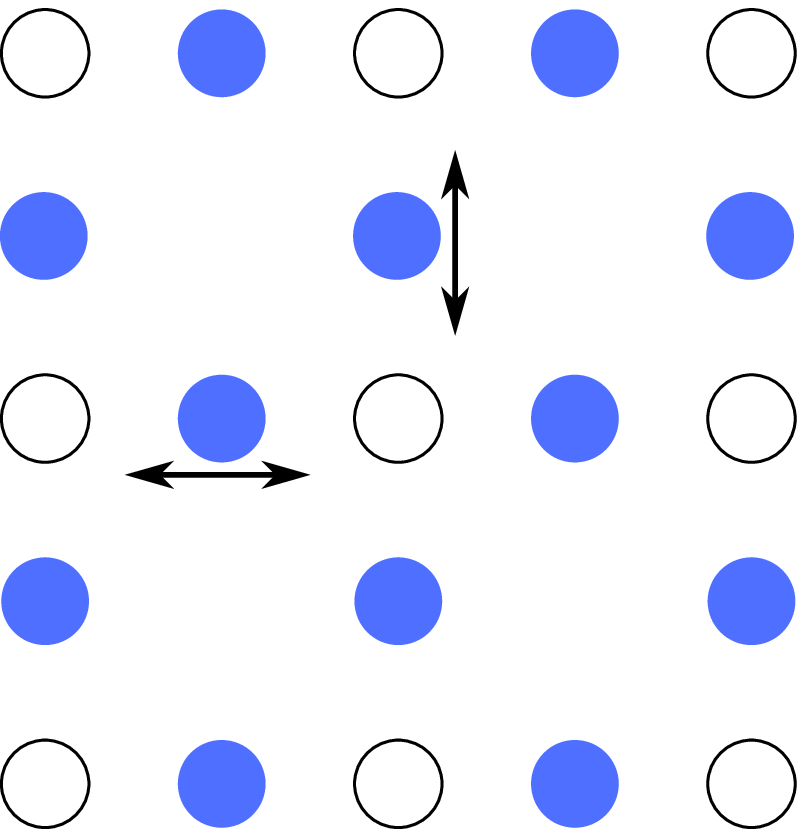}}
  \caption{(color online) Sketch of the crystal structures dis-
cussed in this work: (a) 1D chain, and (b) 2D plane, consist-
ing of light atoms (filled circles) intercalated between heavy
atoms (empty circles). In the absence of carriers, the ionic
potential of a light atom is a simple harmonic well. In the
presence of a carrier, the ionic potential of the light atom
hosting it remains an even function of its longitudinal dis-
placement, so the linear e-ph coupling vanishes. In suitable
conditions the effective ionic potential becomes a double well. 
(Reproduced from Ref.~\onlinecite{double_well})
  \label{fig:example_systems}}
\end{figure}

The model is relevant for crystals whose structure is such that a
sublattice of light ions is symmetrically intercalated with one of
much heavier ions; the latter are assumed to be immobile. Moreover,
charge transport occurs on the sublattice of the light ions. An
example is the one-dimensional intercalated chain shown in
Fig.~\ref{fig:example_systems}(a). Another example is a
two-dimensional CuO layer, sketched in
Fig.~\ref{fig:example_systems}(b), where the doping holes move on the
light oxygen ions placed in between the heavy copper ions. In such
structures, because in equilibrium each light ions is symmetrically
placed between two immobile heavy ions, the potential felt by a
carrier located on a light ion must be an even function of that ion's
longitudinal displacement from equilibrium, i.e. the first derivative
of the local potential must vanish. As a result, the linear
electron-phonon coupling is zero by symmetry, and one needs to
consider the quadratic coupling. This is what the double-well el-ph
coupling model does.

Starting from the single-polaron Hamiltonian describing double-well
el-ph coupling, introduced in Ref.~\onlinecite{double_well}, we add
the appropriate terms for the many-electron problem to obtain
\begin{multline}
  \Ham = \hat T + \Omega \sum_{i} b_{i}^\dg b_{i}^\pd
  + U\sum_i \hat n_{i\uparrow} \hat n_{i\downarrow}  \\
+  g_2 \sum_{i\sigma} c_{i\sigma}^\dg c_{i\sigma}^\pd \left( b_i^\dg + b_i^\pd\right)^2
+ \sum_{i} g_4^{n_i} \left(
  b_i^\dg + b_i^\pd \right)^4.
\label{eq:H}
\end{multline}
Here, $c_{i\sigma}$ and $b_i$ are annihilation operators for a
spin-$\sigma$ carrier at site $i$, and a phonon at site $i$.  $\hat T$
describes hopping of free carriers on the sublattice of light ions in
an intercalated lattice like that sketched in
Fig.~\ref{fig:example_systems}.  For simplicity, we consider
nearest-neighbor hopping only, $\hat T = -t\sum_{\langle
  i,j\rangle,\sigma} c_{i\sigma}^\dg c_{j\sigma}^\pd + h.c.$, although
our method can also treat longer-range finite
hopping.\cite{mirko_efficient} The next two terms describe a single
branch of dispersionless optical phonons with energy $\Omega$, and the
Hubbard on-site Coulomb repulsion with strength $U$.  The last two
terms describe the el-ph coupling in the double-well model. As
mentioned, in lattices like that sketched in
Fig.~\ref{fig:example_systems}, the coupling depends only on even
powers of the light-ion displacement $\delta \hat x_i \propto b_i^\dg
+ b_i$ (the heavy ions are assumed to be immobile).  As a result, the
lowest order el-ph coupling is the quadratic term whose characteristic
energy $g_2$ can have either sign, depending on modeling details.  As
discussed at length in Ref.~\onlinecite{double_well}, the interesting
physics occurs when $g_2 <0$ so that the el-ph coupling ``softens''
the lattice potential. For sufficiently negative $g_2$ this renders
the lattice locally unstable in the harmonic approximation and
requires the inclusion of quartic terms in the lattice potential. For
consistency, one should then also include quartic terms in the el-ph
coupling.  As detailed in Ref. \onlinecite{double_well}, under
reasonable assumptions the quartic lattice terms can be combined with
the quartic el-ph coupling term on sites hosting a carrier and ignored
on all other sites. Because the resulting quartic term contains
contributions from both the lattice potential and from the el-ph
interaction, it should not be assumed to be linear in the carrier
number, unlike the quadratic term which arises purely from el-ph
coupling. Instead, we use the general form
\begin{equation*}
  g_4^{(n_i)} = g_4\cdot \begin{cases} 0, & \mbox{ if }n_i = 0 \\
                         1, & \mbox{ if } n_i = 1 \\
                         \alpha, & \mbox{ if } n_i = 2
             \end{cases}
\end{equation*}
where $n_i=\sum_{\sigma}^{}c^\dg_{i\sigma}c_{i\sigma}$ is the number
of carriers on site $i$, and $\alpha$
is a constant between $1$ and $2$. Setting $\alpha = 2$ assumes that
quartic lattice effects are negligible compared to the quartic el-ph
terms, whereas $\alpha = 1$ is the opposite extreme.
For the remainder of this article we set $\alpha = 1$, so that $g_4^{(1)} =
g_4^{(2)} = g_4$. This case leads to stronger coupling, since a
lower $g_4$ results in deeper wells that are further
apart,\cite{double_well} and thus represents the parameter regime we
are interested in. Physically, this describes the situation where the
quartic lattice terms are much larger than the quartic el-ph coupling;
however they are still negligible compared to the quadratic lattice
terms and therefore can be ignored at sites without a carrier.

\section{Formalism}\label{sec:formalism}

We compute the bipolaron binding energy and effective mass using the
momentum average (MA) approximation.\cite{MA_berciu, MA_goodvin,
  adolphs_nonlinear, double_well} Since we are interested in
strongly-bound bipolarons which have a large probability of having both
carriers on the same site, the version of MA used here is the
variational approximation that discounts states where the two carriers
occupy different sites. This results in an analytic expression of
the two-particle Green's function which is used to efficiently
explore the whole parameter space. The accuracy of this flavor of MA
is verified  by performing
exact diagonalization in a much larger variational subspace (details
are provided below). In the regime of interest the agreement is very
favorable, showing that the effort required to perform the analytical calculation for
a flavor of MA describing a bigger variational space is
not warranted.

\subsection{Momentum Average approximation}
We  define states with both carriers at
the same site, $\ket{i} = c_{i\uparrow}^\dg c_{i\downarrow}^\dg \ket{0}$, and
states of given total momentum $\vec{k}$ with both carriers at the
same site,
\begin{equation*}
\ket{\vec{k}} = \frac{1}{\sqrt{N}} \sum_i e^{i\vec{k}\cdot \vec{r}_i }
\ket{i}.
\end{equation*}
The bipolaron dispersion $E_{\text{bp}}(\vec{k})$ is obtained from the
lowest energy pole of the two-particle Green's function
\begin{equation*}
G(\vec{k},\omega) = \braket{\vec{k} | [\omega - \Ham + i\eta]^{-1} |
\vec{k}},
\end{equation*}
where $\eta \rightarrow 0^+$ is a small convergence factor.  The
effective bipolaron mass is $1/m_{\text{bp}} = \partial^2 E_{\text{bp}}
/ \partial k^2|_{k=0}$. Throughout this work we set $\hbar = 1, a=1$.

We split the Hamiltonian into $\Ham = \Ham_0 + \Ham_1$
with $\Ham_0 = \hat T + \Omega \sum_i b_i^\dg b_i$ describing the free
system and $\Ham_1$ containing the interaction terms. We
apply Dyson's identity $\hat G(\omega) = \hat G_0(\omega) + \hat
G(\omega) \Ham_1 \hat G_0(\omega)$ where
\begin{equation*}
\hat G_0(\omega) = \left[\omega - \Ham_0 + i\eta\right]^{-1}
\end{equation*}
is the resolvent of $\Ham_0$ and we also define
\begin{equation*}
G_0(\vec{k},\omega) = \braket{\vec{k} | \hat G_0(\omega) | \vec{k}} =
\frac{1}{N} \sum_{\vec{q}} \frac{1}{\omega +i\eta -
  \epsilon(\vec{k}-\vec{q}) - \epsilon(\vec{q})}
\end{equation*}
as a non-interacting two-particle propagator, where
$\epsilon(\vec{k})$ is the free carrier dispersion. $N\rightarrow \infty$ is
the number of light-ion sites of the lattice.  In 1D, $G_0(\vec{k},\omega)$
equals the momentum-averaged \emph{single-particle} free
propagator in one dimension for an effective hopping integral
$2t\cos(k/2)$, for which an analytic expression is
known.\cite{Economou} In higher dimensions, such propagators can be
calculated as discussed in Ref. \onlinecite{few_particle}.

As mentioned, in a variational sense the MA used here amounts to
neglecting all states where the
carriers are not on the same site. This approximation is justified for
the description of the strongly bound on-site (S0) bipolaron, which is
expected to have most of its weight in the sector where both carriers
are on the same site. Another way to look at this is that the bipolaron ground-state energy
in the strongly-bound case must be well below the non-interacting
two-particle continuum, and the free two-particle
propagator will have vanishingly small off-diagonal matrix
elements at such energies. Ignoring them, the equation of motion  becomes
$ G(\vec{k},\omega) \approx G_0(\vec{k},\omega)  +
\braket{\vec{k} | \hat G(\omega) \Ham_1 | \vec{k}}
	   G_0(\vec{k},\omega)$, and thus:
\begin{widetext}
\begin{equation*}
G(\vec{k},\omega) = G_0(\vec{k},\omega) \left( 1 + \sum_i \frac{e^{ikR_i}}{\sqrt{N}}
\left[(g_2 + 6g_4) F_1(\vec{k},\omega,i) \\ + g_4 F_2(\vec{k},\omega,i) + U F_0(\vec{k},\omega, i)\right]
\right).
\end{equation*}
\end{widetext}
where $F_n(\vec{k},\omega,i) = \braket{\vec{k}| \hat G(\omega)
  b_i^{\dg, 2n} | i}$ is a generalized two-particle propagator.
Equations of motion for the $F_n$ propagators are obtained in the same
way, and read:
\begin{multline}
F_n (\vec{k},\omega,i)= \bar g_0(\omega - 2n\Omega) \Big[
g_4 (2n)^{\bar 4} F_{n-2}(\vec{k},\omega,i) \\ + \big( (2g_2 + 6g_4) (2n)^{\bar 2} +
4g_4 (2n)^{\bar 3} \big) F_{n-1}(\vec{k},\omega,i)
\\ + (8ng_2 + 12ng_4 + 24n^2 g_4 + U) F_n (\vec{k},\omega,i)\\+ (2g_2 + 6g_4 + 8ng_4) F_{n+1} (\vec{k},\omega,i)+ g_4 F_{n+2}(\vec{k},\omega,i)
\Big].\label{eq:eom}
\end{multline}
where we use the shorthand notation $x^{\bar n} = x! / (x-n)!$ and
have introduced the \emph{momentum-averaged} free two-carrier propagator,
\begin{multline*}
  \bar g_0(\omega) := \braket{i | \hat G_0(\omega) | i} =
\frac{1}{N}\sum_{\vec{k}} G_0(\vec{k},\omega) \\=
\frac{1}{N^2}\sum_{\vec{k},\vec{q}} \frac{1}{\omega -
  \epsilon(\vec{k}-\vec{q}) - \epsilon(\vec{q}) + i\eta}.
\end{multline*}
In 1D, $\bar g_0(\omega)$ equals the diagonal element of the free
propagator for a particle in 2D, which can be expressed in terms of
elliptical functions and calculated efficiently.\cite{Economou}
Similar considerations hold in higher dimensions.\cite{few_particle}

The equations of motions are then solved following the procedure
described at length in Refs.~\onlinecite{double_well,
  adolphs_nonlinear}. For consistency, we sketch the main steps here.
First, we introduce vectors $W_n = (F_{2n-1}, F_{2n})^T$ for $n \ge 0$
(the arguments $\vec{k},\omega,i$ of the propagators are not written
explicitly from now on). Note that with this definition, $W_0 =
(F_{-1}, F_0)$, yet $F_{-1}$ is not properly defined. However, the
final result has no dependence on $F_{-1}$, as we show below.  The
equations of motion are then rewritten in terms of  $W_n$ to read $\gamma_n W_n =
\alpha_n W_{n-1} + \beta_n
W_{n+1}$. The matrix elements of the $2\times
2$ matrices $\alpha_n, \beta_n, \gamma_n$, are easily read off Eq.~\eqref{eq:eom}.

Defining
$A_n = [ \gamma_n - \beta_n A_{n+1} ]^{-1} \alpha_n$, the physical
solution of these recurrence equations is $W_n = A_n W_{n-1}$.
Introducing a sufficiently large cut-off $N_c$ where $W_{N_c} = 0$, we
can then compute $A_1$ and have 
$W_1 = A_1 W_0$, i.e.,
\begin{equation*}
\begin{pmatrix}F_1 \\ F_2 \end{pmatrix} =
A_1 \begin{pmatrix} F_{-1} \\ F_0 \end{pmatrix} =
\begin{pmatrix} a_{11} & a_{12} \\ a_{21} & a_{22} \end{pmatrix}
\begin{pmatrix} F_{-1} \\ F_0 \end{pmatrix}.
\end{equation*}
One can easily check that $a_{11} = a_{21} = 0$. Thus, we obtain
$F_1 = a_{12} F_0$ and $F_2 = a_{22} F_0$. Substituting these results back into the
EOM for $G$ we obtain
\begin{multline*}
G(\vec{k},\omega) = G_0(\vec{k},\omega) \Big[1 +\\
 \sum_i \frac{e^{i\vec{k}\vec{R}_i}}{\sqrt{N}}
\left((2g_2 + 6g_4) a_{12} + g_4 a_{22} + U\right) F_0(\vec{k},\omega,i) \Big].
\end{multline*}
Since, by definition, $G(\vec{k},\omega) =  \sum_i \frac{e^{i\vec{k}\vec{R}_i}}{\sqrt{N}}
F_0(\vec{k},\omega,i)$, and given that $a_{12}, a_{22}$ are functions
of $\omega$ only,  we find:
\begin{equation}
G(\vec{k},\omega) = \frac{1}{G_0^{-1}(\vec{k},\omega) -
(2g_2 + 6g_4)a_{12} - g_4 a_{22} - U}
\label{eq:se}.
\end{equation}

Note that the coefficients $a_{12}$ and $a_{22}$
depend on all  parameters of the model, including $U$. As a result, the
position of the lowest pole of Eq.~\eqref{eq:se} is \emph{not}
simply linear in $U$, although this is a good approximation for the
strongly bound bipolaron.

We emphasize that this MA expression becomes exact in two limiting
cases. First, in the atomic limit $t \rightarrow 0$ the free
propagator has no off-diagonal terms and thus no error is introduced
by dropping them from the equations of motion. Second, without el-ph
interactions ($g_n = 0$) the Hamiltonian reduces to the Hubbard model
which is exactly solvable in the two-particle
case.\cite{sawatzky_quasiatomic} In both cases MA gives the exact
solution.

\subsection{Exact Diagonalization}
The results obtained via MA as outlined above are checked against
exact diagonalization results in a bigger variational subspace
designed to describe well the strongly bound S0 bipolaron. Hence,
we only consider states where all the phonons are located on the same
lattice site and at least one of the two electrons is close to this
cloud. The basis states are of the form
\begin{equation*}
\ket{\vec{k}, n, \delta_1, \delta_2} = \sum_i
\frac{e^{i\vec{k}\vec{r}_i}}{\sqrt{N}} b_i^{\dg, n}
c_{i+\delta_1,\uar}^\dg c_{i+\delta_2, \dar}^\dg \ket{0}
\end{equation*}
with the constraint that either $\delta_1$ or $\delta_2$ is below a
certain cut-off.  In addition, a global cut-off $N_c$ is imposed on $n
+ \delta_1 + \delta_2$.  The ground state within the variational space
is then computed using standard eigenvalue techniques.

The main difference between these ED and MA results is that MA discards
contributions from configurations where the carriers are at
different lattice sites. Comparing the two therefore allows us to gauge the
importance of such terms, and to decide
whether
the speed gained from using the analytical MA expressions
counterbalances the loss of accuracy.

\section{Results}\label{sec:results}

\begin{figure}[t]
\begin{center}
\includegraphics[width=0.48\textwidth]{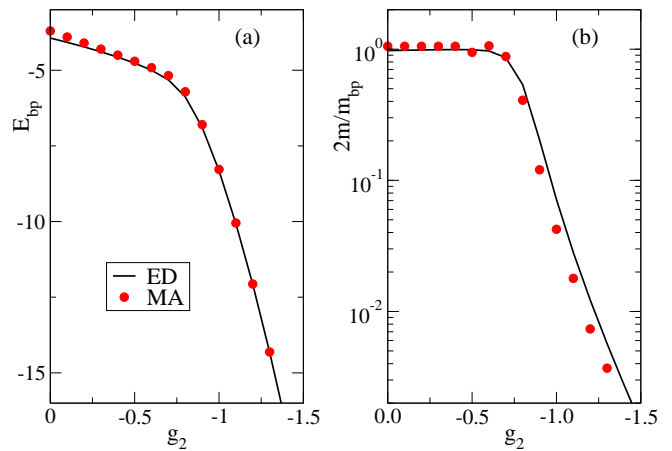}
\caption{(color online)
 (a) Bipolaron ground-state energy, and (b) inverse effective mass
for $t=1, \Omega = 0.5$, $g_4 = 0.1$, computed with ED (solid black line)
and MA (red dots). \label{fig:ED_MA_compare}}
\end{center}
\end{figure}

From now on we focus on the one-dimensional (1D) case, since our previous
work\cite{double_well} suggests that going to higher dimensions leads
to qualitatively similar results.

Before discussing the MA results, we first compare them to
those obtained from ED in the larger variational subspace
discussed above. A typical comparison (for $t=1, \Omega = 0.5$ and $g_4 =
0.1$) is shown in Fig.~\ref{fig:ED_MA_compare}. The left panel shows
the ground state energy and the right panel shows the inverse
effective mass of the bipolaron. In the regime where the bipolaron is
strongly bound, {\em i.e.} where its energy decreases fast and its
effective mass increases sharply as $|g_2|$ increases, we find
excellent agreement for the energy. The masses
also agree reasonably well, but MA systematically \emph{overestimates}
the bipolaron mass. This is a direct result of the more restrictive
nature of the MA approximation: By discarding configurations where the
carriers occupy different sites, the mobility of the bipolaron is
underestimated and thus the effective mass is
overestimated. Nonetheless, this error is not very large, and only
means that the bipolarons in the double well model are even lighter
than calculated by MA. Due to similarly good agreement in all cases we
verified, for the remainder of this work we only discuss results
obtained with the more efficient MA method.

We emphasize that our approximation for computing the Green's function is
only valid in the regime of strong binding and does not describe
correctly the physics at weak coupling. Since neither MA nor ED, as
implemented here, allow for the formation of two phonon clouds,
neither describes the formation of a weakly bound S1
bipolaron (where polarons form on neighboring sites and interact with
each other's clouds via virtual hoppings), nor the dissociation into two
 polarons as the coupling is further
decreased.\cite{bonca_bp,macridin} Accuracy in these parameter
regimes can be improved by applying more sophisticated -- yet much
more tedious -- versions of MA or ED for suitably expanded variational
spaces. For the purpose of this work,
however, we want to focus on the strong-coupling regime, where our
results are accurate.

\begin{figure}[t]
\includegraphics[width=0.48\textwidth]{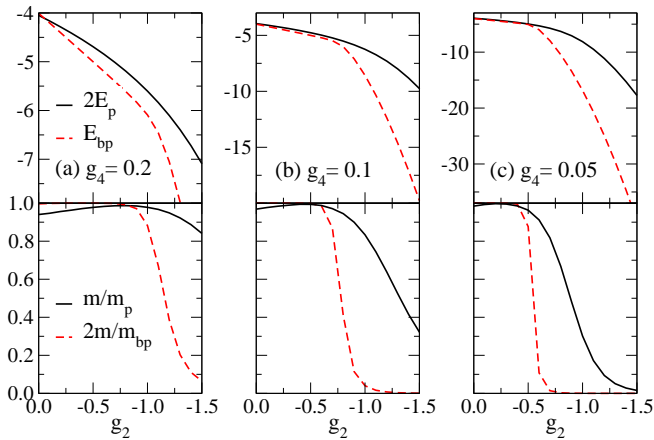}
\caption{(color online) Ground-state properties (total energy and inverse mass)
of the S0 bipolaron and two independent
polarons for $t=1, \Omega=0.5$ and $g_4=0.2$, $0.1$, and $0.05$ for a),
b), and c), respectively. For all panels, $U = 0$.
\label{fig:comparison1}}
\end{figure}

\begin{figure}[b]
\includegraphics[width=0.48\textwidth]{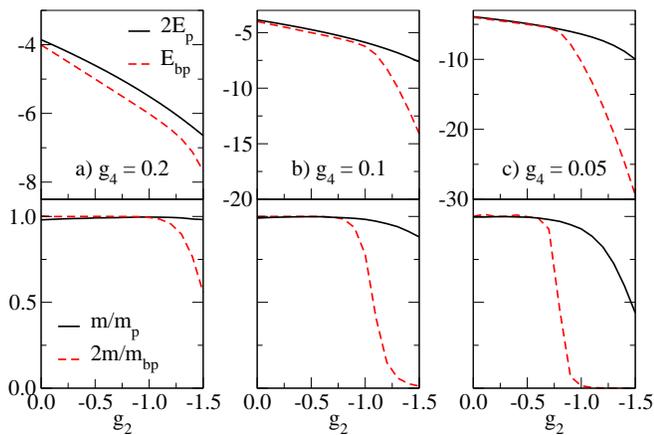}
\caption{(color online) Ground-state properties (total energy and inverse mass)
of the S0 bipolaron and two independent
polarons for $t=1, \Omega=2$ and $g_4=0.2$, $0.1$, and $0.05$ for a), b),
and c), respectively. For all panels, $U = 0$.
\label{fig:comparison2}}
\end{figure}

We show the ground-state properties of the bipolaron compared to those
of two single polarons in Figs.~\ref{fig:comparison1},
\ref{fig:comparison2} for two different values of $\Omega$. In all
those panels, we have set $U = 0$ for simplicity; the role of finite
$U$ will be discussed at the end of this section.

\begin{figure}[b]
\includegraphics[width=0.48\textwidth]{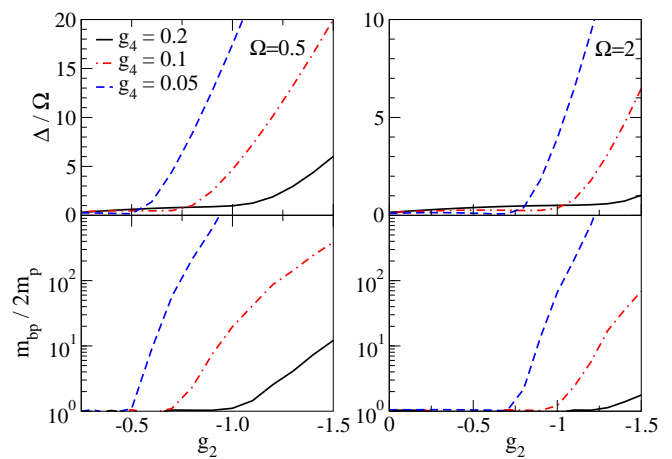}
\caption{(color online) Binding energy $\Delta$ and effective-mass
  ratio $m_{\rm bp} / 2m_{\rm p}$ of the bipolarons for $\Omega = 0.5$
  and $\Omega = 2$ at different values of
  $g_4$. \label{fig:all_combined}}
\end{figure}

The ground state energy of the bipolaron behaves qualitatively
similar for all values of $\Omega$ and $g_4$ in that it shows a kink
at some $g_2$ where the slope becomes steeper. This signifies the
onset of the strong-coupling regime where the bipolaron
energy is well below the energy of two independent polarons,
consistent with a  strongly bound bipolaron. At weaker
coupling the results are not accurate since -- as explained above --
our version of MA cannot describe the dissociation of the bipolaron.

Consider now the behavior of the effective masses. For all parameters
considered here, we see that the single polaron
mass $m_p$ starts out slightly above the free electron mass $m$, then
decreases until it is almost as light as the free electron, before
increasing again. This turnaround in the polaron mass is due to
partial cancellation effects of the quadratic and quartic el-ph coupling
terms, as discussed in Ref. \onlinecite{double_well}. We observe that
when the \emph{bipolaron} is already quite strongly
bound, the single polaron can still be very light. Empirically, we find
that in the strong-coupling regime $m_{\rm p} \sim m
\exp(-\gamma \Delta_{\rm p} / \Omega)$ where $\Delta_{\rm p}= -2t -
E_{\rm p}$ is the single-polaron binding energy and $\gamma$ is a
small numerical prefactor. This behavior is also found in the Holstein
model\cite{holstein} in the strong-coupling limit, where $\gamma=1,
\Delta_{\rm p}=-g^2/\Omega$. The prefactor $\gamma$ can be much smaller in the
double well model because of the nature of the ionic potential. This
was explained in detail in Ref.  \onlinecite{double_well}, and will be
discussed in the context of bipolarons later in this section.

The bipolaron effective mass fluctuates around the value of $2m$ in
the weak coupling regime. As explained above, here our method does not
describe two independent polarons, but two independent free electrons
whose effective mass should just be $2m$. However, the two-particle
spectral function in this case does not have a low-energy
quasi-particle peak. Instead, it has a continuum spanning the allowed
two-particle continuum. 

These issues disappear at stronger coupling where a strongly bound
bipolaron forms and MA becomes accurate. The figures show that here
the bipolaron quickly gains mass with increased coupling strength
$|g_2|$, and that this increase is stronger the \emph{smaller} $g_4$
is.  Note that a smaller $g_4$ actually means stronger coupling,
because the wells are deeper and further apart.\cite{double_well}

The same data is displayed in a different way in
Fig.~\ref{fig:all_combined}, where we show the magnitude of the bipolaron binding
energy $\Delta = 2E_{\rm p} - E_{\rm bp}$ and the ratio of bipolaron
to single-polaron masses, $m_{\rm bp} / 2m_{\rm p}$. The strongly bound
bipolaron regime (where the results are accurate) is reached when these
quantities vary fast with  $g_2$. In particular, the results for
$m_{\rm bp} / 2m_{\rm p}$ show that here the bipolaron mass increases
much more quickly than the polaron mass. This
is not surprising for models like this, where the phonons modulate the
on-site energy of the carrier. At strong coupling the
results can be understood
starting from the atomic limit $t = 0$, treating hopping as a
perturbation. Since both carriers must
hop in order for the bipolaron to move, one expects that $m_{\rm
  bp}/m \propto (m_{\rm
  p} / m)^2$; indeed, we find this relation to be valid for a wide range of
parameters for our model.

Although in this regime the bipolaron quickly gains mass, there are
parameter ranges where its mass is still rather light
while the bipolaron is strongly bound. Examples of such
parameters are given in Table~\ref{tab:examples}.  We note that
qualifiers such as ''strongly bound`` and ''light`` are subjective.
In our case, we take the bipolaron as strongly bound when the binding
energy $\Delta/\Omega >1$ and
the ratio $m_{\rm bp}/2m < 10-20$, consistent with other
references.\cite{hague_light_bp,extended_HH}

\begin{table}
\caption{Some example values of the bipolaron binding energy and effective mass.
\label{tab:examples}}
\begin{ruledtabular}
\begin{tabular}{ccccc}
$\Omega$ & $g_4$ & $|g_2|$ & $\Delta/t$ & $m^{**} / 2m$ \\\hline
$0.5$ & $0.1$ & 0.9 & 1.25 &	8.3 \\
      & $0.2$ & 1.3 & 1.48 & 4.4 \\\hline
$2$   & $0.1$ & 1.3 & 3.11 & 5.9 \\
	  & $0.2$ & 1.5 & 1.03 & 1.8
\end{tabular}
\end{ruledtabular}
\end{table}

Light but strongly bound bipolarons were previously found for
long-range el-ph coupling.\cite{extended_HH} The explanation is
that in such models, carriers induce a spatially extended lattice
deformation, not one that is located in the immediate vicinity of the
carrier as is the case at strong coupling in local el-ph coupling
models. Because of their extended nature, the overlap between clouds
displaced by one lattice site (which controls the effective hopping)
remains rather large, meaning that the polarons and bipolarons remain
rather light in such models.

\begin{figure}[t]
\begin{center}
\includegraphics[width=0.48\textwidth]{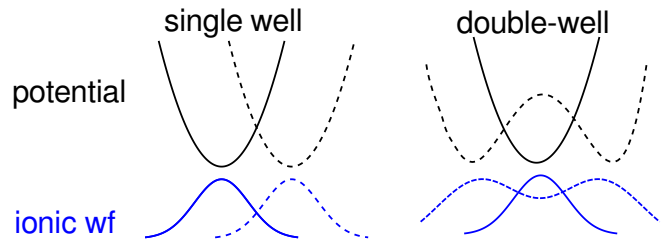}
\caption{(color online)
Ionic potential (above) and ionic ground-state wavefunction (below)
in the single-well and double-well models. Solid lines correspond
to the situation without an additional carrier, dashed lines to
the situation with an additional carrier.
 \label{fig:potential_overlaps}}
\end{center}
\end{figure}

Even though it is due to a local el-ph coupling, the mechanism
resulting in light bipolarons in our model is qualitatively similar,
as illustrated in Fig.~\ref{fig:potential_overlaps}.
In the linear Holstein model the
  effect of an additional carrier added to a lattice site is to shift
  the equilibrium position of the ionic potential. The ionic
  wavefunctions corresponding to an empty and an occupied site
  therefore have only small overlap, which strongly reduces the
  effective carrier hopping.  In the double-well model, in contrast, the
  ionic wavefunction for the doubly-occupied site has
  appreciable overlap with the ionic wavefunction for an
  empty or a singly-occupied site and thus does not reduce the
  effective hopping as much.

\begin{figure}[b]
\begin{center}
  \includegraphics[width=0.48\textwidth]{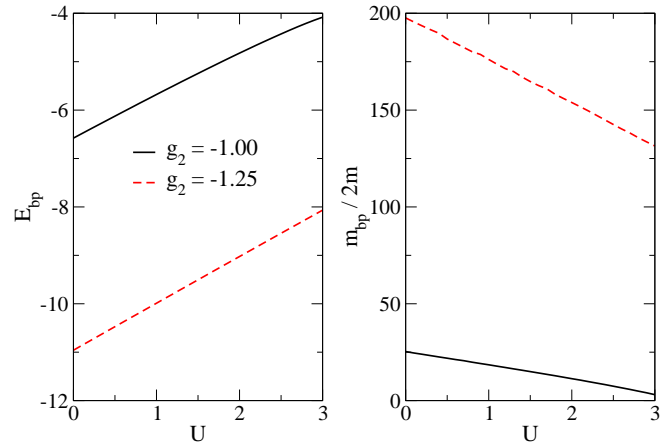}
  \caption{\label{fig:7}(color online) Bipolaron energy (left) and
    effective mass (right) as a
    function of the Hubbard $U$, for $t=1, \Omega=0.4$ and
    $g_4=0.1$. Similar results are found for other parameters
    if $U$ is not large enough to lead to bipolaron dissociation.}
\end{center}
\end{figure}

We conclude with a brief discussion of the effects of a finite,
repulsive $U$. For a very strongly coupled S0 bipolaron, most of the
weight is in states with both carriers on the same site. In 
this regime, the binding energy decreases (nearly) linearly with $U$,
$\Delta_{\rm bp}(U) \approx \Delta_{\rm bp}(U=0) - U$. However, increasing $U$
increases the energy cost of the S0 state and thus encourages
hybridization with off-site states, which results in an overall
smaller effective mass. We show results for the bipolaron energy and
effective mass as a function of $U$ in Fig.~\ref{fig:7}.  We stay
within the regime $U < \Delta_{\rm bp}$ where the bipolaron remains
strongly bound. As predicted, the energy of the bipolaron increases
linearly with $U$, which in turn means that the binding energy
$\Delta_{\rm bp}$ decreases linearly with $U$. The effective mass also
decreases (approximately) linearly with $U$. This can be demonstrated
for the strong coupling limit via second order perturbation theory in
the hopping. Following along the lines in Refs.~\onlinecite{bonca_bp, macridin},
the effective hopping of the S0 bipolaron is of the form
\begin{equation*}
  m^{-1}_{\rm bp} \propto t_{\rm eff} \sim \frac{-t^2 e^{-\gamma \Delta / \Omega}}{2E_p - U}
\end{equation*}
for some constant $\gamma$. We see that the mass itself decreases
linearly with $U$, with a steeper slope the larger the effective
mass at $U = 0$.

In essence, provided that it is
not large enough to break the bonding, a finite $U$ does not change
the overall picture and merely tunes the balance between the bipolaron binding
energy and its effective mass.

\section{Conclusions and Outlook}\label{sec:conclusion}
In conclusion, we have investigated the bipolaron ground-state
properties in the dilute limit of the double-well el-ph coupling model
at strong coupling. We have demonstrated that due to the particular
nature of the carrier-induced ionic potential, the double-well
bipolaron can be strongly bound while remaining light compared to the
bipolaron in the Hubbard-Holstein model. This suggests a new route to
stabilizing such bipolarons, in addition to
previously discussed mechanisms based on long-range el-ph coupling or
special lattice geometries. We expect that a combination of these
mechanisms will lead to even lighter bipolarons.

In this work, we have used and validated a simple extension of the
Momentum Average approximation to the two-carrier case. While this
generalization is appropriate to describe a strongly bound S0
bipolaron, it cannot describe the off-site (S1) bipolaron that forms
at larger Hubbard repulsion $U$, or the unbinding of the bipolaron at
even larger $U$.  A more sophisticated version of MA, currently under
development, will give us insight into the full phase diagram of the
double-well model.

\acknowledgements
Financial support from NSERC and the UBC  Four Year Doctoral Fellowship
program are acknowledged.

\end{document}